\documentclass[preprint,showpacs,preprintnumbers,amsmath,amssymb,endfloats*]{revtex4}
\usepackage{graphicx}

\newcommand{\bes}{\begin{eqnarray}}
\newcommand{\ees}{\end{eqnarray}}

\begin{document}

\thispagestyle{empty}
\title{Investigation of the Casimir force between metal
and semiconductor test bodies
}
\author{F.~Chen,${}^{1}$ U.~Mohideen,${}^{1}$
  G.~L.~Klimchitskaya,${}^{2}$
  and
V.~M.~Mostepanenko${}^{3}$
}

\affiliation{${}^{1}$Department of Physics, University of California,
Riverside, California 92521, USA \\
${}^{2}$North-West Technical University, Millionnaya St. 5, St.Petersburg,
191065, Russia\\
${}^{3}$Noncommercial Partnership ``Scientific Instruments'', 
Tverskaya St.{\ }11, Moscow, 103905, Russia 
}

\begin{abstract}
The measurement of the Casimir force between a large gold coated sphere
and single crystal silicon plate is performed with an atomic
force microscope. A rigorous statistical comparison of data with theory is
done, without use of the concept of root-mean-square deviation, and 
excellent agreement is obtained. The Casimir force between metal and
semiconductor is demonstrated to be qualitatively different than
between two similar or dissimilar metals which opens new opportunities
for applications in nanotechnology.
\end{abstract}


\maketitle

In this paper we present the results of the experimental and theoretical
investigation of the Casimir force acting between a gold coated sphere
and a single crystal silicon plate. The Casimir
force is determined by the alteration of zero-point oscillations
of the electromagnetic field due to the presence of material boundaries
(see the original paper \cite{1} and monographs \cite{2,3,4}).
The Casimir force is in fact the limiting case of the van der Waals force
when the separation between the test bodies becomes large enough  
for retardation to be included. Historically, most
measurements of the Casimir force were performed between dielectrics
(see Ref.~\cite{5} for review), and anomalous behavior in silicon has been
reported \cite{5a}. In the last few years many
precise experiments using metallic test bodies have been done and the
results were compared with theory taking into account different
corrections to the Casimir force \cite{5,6,7,10,11,12,13}.
The obtained results have been found to be of prime importance in the
physics of nano- and micromechanical systems \cite{14} and for
testing predictions of extra-dimensional models and other theoretical
schemes beyond the standard model \cite{13}.

To gain a better insight of the role of the Casimir effect in
nanotechnology, it is important to understand the effect of semiconductor 
test bodies.  These materials are central to the fabrication and 
design of nano- and microdevices and provide a wide variety of
electrical properties which may influence the Casimir force.
Until now, however, no precise experiments on the Casimir effect
with semiconductor bodies have been performed (see the discussion
on the importance of this subject in Ref.~\cite{15}).
Below we demonstrate that the ratio of the Casimir forces between
Au and Si test bodies to that from Au-Au decreases with the increase
of separation. This is qualitatively different from the case
when Si is replaced with some metal of lower conductivity than
Au where the same ratio is practically constant or increases with
separation.
Another important point of this paper is the comparison between the
measurement data and theory without the use of the concept of the
root-mean-square deviation widely employed in previous experiments
on the Casimir effect. As was shown in Ref.~\cite{15a}, this approach
may be inadequate when the measured force rapidly changes with
separation distance, though no better approach was suggested.

One test body is a sphere attached to a cantilever of an
atomic force microscope (AFM).
The sphere is coated with an Au layer of 105\,nm
thickness. The diameter of the sphere was measured using a scanning
electron microscope to be
$2R=(202.6\pm 0.3)\,\mu$m.
The other test body is $5\times 10\,\mbox{mm}^2$
single crystal silicon Si$<100>$ plate.
The nominal resistivity of the Si plate was 
(0.01--0.001)\,$\Omega\,\mbox{cm}$.
Using the four probe technique we measured its resistivity to be
$\rho=0.0035\,\Omega\,\mbox{cm}$. 
Note that for all frequencies contributing to the Casimir force
this Si plate, unlike metals, has a large absorption typical of 
semiconductors (metallic resistivities are
usually two or three orders of magnitude lower).
The Casimir force acting
between the Au sphere and Si plate was measured by means of the improved
setup previously used in Ref.~\cite{10} for two Au test bodies.
The main improvements and innovations implemented in this experiment
are: We now use much higher vacuum
$2\times 10^{-7}\,$Torr (instead of $3\times 10^{-2}\,$Torr in
Ref.~\cite{10}) to maintain the chemical purity of the Si surface
wich oxidizes rapidly to SiO${}_2$. In addition, this vacuum 
system is oil free, consisting of oil free mechanical
pumps, turbo pumps and ion pumps to prevent contamination.
To reduce the influence of mechanical noise during data
acquisition,
only the ion pump is used to maintain the lowest pressures.

A special passivation procedure is used to prepare the Si surface.
First nanostrip (combination of H${}_2$O${}_2$ and  H${}_2$SO${}_4$)
is used to clean the surface of organics and other contaminants.
This oxidizes the surface. Then we use 49\% HF to etch SiO${}_2$
and hydrogen to terminate the surface. 
The termination is stable for more than two weeks under the vacuum
conditions described above \cite{si}.
The bottom of the Si plate is
coated with about 100\,nm of Au and used for the electrical contact.
It was checked to be ohmic in nature. 
Above steps were necessary
to keep the residual potential difference low, constant and
independent of separation distance.

The next improvement is the reduction of the uncertainty in the
determination of absolute separation distances $z$
down to $\Delta z=0.8\,$nm (in comparison with
1\,nm in Ref.~\cite{10}). To achieve this aim, here we use a piezo
capable of travelling a distance of 6\,$\mu$m from initial
separation to contact of the two surfaces (previously \cite{10}
the movement of the plate to large separations was done mechanically 
and the piezo movement was used only at short separations of less 
than 2 $\mu$m). All 6\,$\mu$m of piezo movement
are calibrated interferometrically. 
As a result, the error in the piezo calibration practically does
not contribute to $\Delta z$.
Then different DC
voltages between +0.2\,V to --0.4\,V were applied to the plate and 
the electric force was measured. The electric force measurement with
each voltage was repeated 5 times and the average value was used
to fit the exact electrostatic
force-distance relation \cite{10} to determine the 
separation on contact of the two surfaces $z_0$.
The resulting value, which is not zero
due to the roughness of surfaces, is $z_0=(32.1\pm 0.8)\,$nm.
The error in $z_0$ completely determines the error $\Delta z$
of all measured absolute separations $z$. The values of $z$
are found independently, without fitting to the theoretical 
expression for the Casimir force.

The same procedure also allowed an independent determination of the
residual potential difference $V_0$ at different separations.
The $V_0$ was determined to be 
$V_0=(-0.114\pm0.002)\,$V and
independent of the 
separation. This allowed us to confirm the absence of any 
contamination of the Si surface and the absence of localized charges
(the presence of localized charges would lead to dipole and other 
multipolar electrostatic fields, resulting in a $V_0$ which 
varies with distance). The high conductivity of the Si plate used is
important in preventing the formation of such charges.

Finally the Casimir force between the sphere and the plate as a
function of distance is measured. The sphere is kept grounded
while a compensating voltage corresponding to $V_0$ is applied
to the plate to cancel the residual electrostatic force. 
The distance
was varied from large to short separations by applying 
a continuous voltage to
the piezo. The force data $F^{\rm expt}(z_i)$ were collected at
equal time intervals corresponding to
separations $z_i$ having
a step size of 0.17\,nm. This measurement was repeated for $n=65$
times.

We now turn to a determination of experimental errors and precision.
First the experimental points were analysed for the presence
of so called ``outlying'' results using the statistical criteria
of Ref.~\cite{16}. It was found that none of the $n=65$ sets of measurements
are outlying and can be used in error analysis. To find the random
error the mean values of the measured force over all sets of
measurements ${\bar{F}}^{\rm expt}(z_i)$
are calculated at all points $z_i$
($1\leq i\leq 3164$). The mean experimental force as a function of 
separation for the distance range 62.33\,nm to 600.04\,nm is shown in Fig.~1.
An estimate for the variance of this mean $s_{\bar{F}}(z_i)$
is  not
uniform, i.e., changes randomly when the separation changes less than
$\Delta z=0.8\,$nm. In this case the best estimate for a variance is
calculated by a statistical procedure \cite{17} 
in the theory of repeated measurements 
(see Ref.~\cite{19a} for details).  
Then the variance is approximately
the same for all $z_i$ and equal to $s_{\bar{F}}=1.5\,$pN.

Using the Student's $t$-distribution with a number of
degrees of freedom $f=n-1=64$ and choosing $\beta=0.95$ (hypothesis
is true at 95\% confidence) we obtain $p=(1+\beta)/2=0.975$ and
$t_p(f)=2.00$. Then for the confidence interval it follows
$|{\bar{F}}^{\rm expt}(z)-F(z)|\leq
\Delta^{\!{\rm rand}}{{F}}^{\rm expt}\equiv
s_{\bar{F}}t_p(f)\approx 3.0\,\mbox{pN}$.
Here $F(z)$ is the true value of the Casimir force at a separation $z$
(this exact value can only be approached with complete
knowledge of all possible corrections)
and $\Delta^{\!{\rm rand}}{{F}}^{\rm expt}$ is the random absolute 
error of force measurements in the present experiment. It is almost two
times smaller than the random error in the experiment of
Ref.~\cite{10} with two gold test bodies.

The systematic errors are the same as in the experiment
with two gold bodies (see the second paper in Ref.~\cite{10}).
They  are given by the error
in force calibration 
($0.82\,$pN),
by the noise when the calibration voltage is applied to the cantilever
($0.55\,$pN), by the instrumental
sensitivity ($0.31\,$pN), and 
by the restrictions on computer resolution of 
data ($0.12\,$pN).
The combined systematic error in Ref.~\cite{10} was, however,
overestimated. To obtain the best estimate for it, the difference
between the experimental and true force values at each separation
is assumed to be distributed uniformly.
The resulting systematic error 
$\Delta^{\!{\rm syst}}F^{\rm expt}\approx 1.17\,$pN
at 95\% confidence is given by
the composition of $N$ uniform distributions \cite{16}
(in contrast with 2.7\,pN obtained in Ref.~\cite{10}).
The total experimental error of the Casimir force 
$\Delta^{\!{\rm tot}}F^{\rm expt}
\approx 3.33\,\mbox{pN}$ at 95\% confidence
is obtained from Ref.~\cite{16}  by 
combining the above random and systematic errors.
In Fig.~2 the relative error 
$\delta^{\rm expt}=\Delta^{\!{\rm tot}}F^{\rm expt}/{\bar{F}}^{\rm expt}$
is given by the solid curve as a function of separation. 
It is equal to only 0.87\% at the shortest separation
of 62.33\,nm and increases with an increase of a separation.

For separations used here the thermal corrections 
at $T=300\,$K are not
significant. 
As noted in Ref.~\cite{24}, in this case
one can use the Lifshitz formula at zero temperature
\cite{21b}

\begin{eqnarray}
&&F_c(z)=\frac{\hbar R}{2\pi}
\int_0^{\infty}k_{\bot}dk_{\bot}
\int_0^{\infty}d\xi
\left\{\ln\left[1-r_{\|}^{(1)}r_{\|}^{(2)}
e^{-2zq}\right]\right.
\nonumber \\
&&
\phantom{aaaa}\left.
+\ln\left[1-r_{\bot}^{(1)}r_{\bot}^{(2)}
e^{-2zq}\right]\right\}.
\label{n5}
\end{eqnarray}
\noindent
The reflection coefficients for two independent polarizations are given by
\begin{equation}
r_{\|}^{(p)}=
\frac{\varepsilon^{(p)}(i\xi)q-k^{(p)}}{\varepsilon^{(p)}(i\xi)q+k^{(p)}},
\quad
r_{\bot}^{(p)}=
\frac{k^{(p)}-q}{k^{(p)}+q},
\label{n6}
\end{equation}
\noindent
where
$q^2\equiv k_{\bot}^2+{\xi^2}/{c^2}$,
${k^{(p)}}^2\equiv k_{\bot}^2+\varepsilon^{(p)}(i\xi){\xi^2}/{c^2}$,
and $\varepsilon^{(p)}(\omega)$ is the dielectric permittivity of
gold ($p=1$) and silicon ($p=2$).

$\varepsilon^{(1)}(i\xi)$
was found \cite{5} by means of the dispersion relation from 
the imaginary part of $\varepsilon^{(1)}(\omega)$ obtained using
the complex refractive index from tables \cite{20}. The same procedure
was used for single crystal Si. Since the optical properties of Si
at low frequencies depend on the concentration of charge carriers,
the tabulated data in Ref.~\cite{20}, obtained for a sample of high
resistivity $\rho_0=1000\,\Omega\,$cm, should be adapted for the
silicon plate used in our experiment with a resistivity
 $\rho=0.0035\,\Omega\,$cm.
This is achieved \cite{20} by adding the imaginary
part of the Drude dielectric function to the imaginary part of the 
dielectric permittivity obtained using the data from tables. In doing
so the plasma frequency for Si at a resistivity $\rho$ is found from
$\omega_p^{Si}=2\sqrt{\pi}/\sqrt{\varepsilon_0\rho\tau^{Si}}=
6.37\times 10^{14}\,$rad/s, where $\varepsilon_0$ is the dielectric 
permittivity of vacuum, 
$\tau^{Si}=1/\gamma^{Si}=10^{-13}\,$s \cite{20} is the Si relaxation time,
and $\gamma^{Si}$ is the relaxation parameter
(note that change of $\omega_p$ even by a factor of 1.5 leads to less
than a 1\% change in the Casimir force magnitudes within the
entire separation region).
Within our range of characteristic frequences there are only
negligible differences in the values of
$\varepsilon^{(1)}(i\xi)$ found for the sample of
resistivity $\rho$, used in this experiment, as compared to Si
with much higher resistivity $\rho_0$, as in the tables \cite{20}.
Thus, the relatively high conductivity of our
Si plate plays an important role in avoiding charging but,
at the same time, the sample demonstrates the typical
semiconductor frequency-dependence of $\varepsilon(i\xi)$ 
within the frequency  range contributing
to the force.

For comparison of the theoretical results with the experiment, one
should take into account the surface roughness corrections. The topography
of Au coating on the sphere and of the Si plate was measured using
an AFM. It was found that roughness is mostly 
represented by the stochastically distributed distortions with the
typical heights 11--20\,nm on the sphere and 0.3--0.6\,nm on the Si plate.
There are also rare pointlike peaks on the sphere with 
the heights up to 25\,nm. Denoting by $v_k^{(p)}$
the fractions of the surface area with roughness height
$h_k^{(p)}$ ($p=1$ for a sphere and $p=2$ for a plate), one can find
the zero roughness levels $H_0^{(1)}\approx 15.35\,$nm,
$H_0^{(2)}\approx 0.545\,$nm. 
In the additive approach
 the theoretical Casimir force including both finite
conductivity and surface roughness corrections can be calculated as
\cite{5,7,10,13}
\begin{equation}
F^{\rm theor}(z_i)=\sum\limits_{k,j}
v_k^{(1)}v_j^{(2)}
F_c({\tilde z}_i),
\label{n9}
\end{equation}
\noindent
where ${\tilde z}_i=z_i+H_0^{(1)}+H_0^{(2)}-h_k^{(1)}-h_j^{(2)}$,
and the values of $F_c$ are obtained from Eq.~(\ref{n5}). 
As it was 
demonstrated in Refs.~\cite{10,13}, for such values of roughness the
diffraction-type contributions \cite{21,22} are negligible. 

The two main errors in the theoretical Casimir force
$\delta_{m}^{\rm theor}=\Delta^{\!(m)}F^{\rm theor}/|F^{\rm theor}|$
are due to the use of the proximity force theorem ($m=1$) and due to
sample to sample variations of the optical data ($m=2$).
As is concluded in Ref.~\cite{10}, $\delta_{1}^{\rm theor}<z/R$
and $\delta_{2}^{\rm theor}\approx 0.5$\% (the other errors
contained in the theoretical force 
due to the influence of patch potentials, spatial dispersion and
finite size of the plate
were shown \cite{10,19a} to be much smaller).
In the absence of exact information, both random quantities are
assumed to be distributed uniformly 
(i.e., can take any value with equal probability 
within the fixed intervals determined by the
respective absolute errors; this assumption
is the most conservative because the use
of any other probability distribution rule leads to a smaller 
combined error). For this reason
the resulting error $\delta_{0}^{\rm theor}$ at 95\% confidence
can be found once more from Ref.~\cite{16}.
Another type of error
in the theoretical Casimir force arises when one 
substitutes into Eqs.~(\ref{n5}), (\ref{n9}) the experimental data for
separations $z_i$ and sphere radius $R$. It is given by \cite{23}
$\delta_{3}^{\rm theor}\approx \Delta R/R+3\Delta z/z$ 
(here we do not use the additional
fit \cite{10} in order to decrease $\Delta z$ because the comparison
between theory and experiment is not based on the 
root-mean-square deviation). 

To determine the total theoretical error of the Casimir force computations
$\delta^{\!{\rm theor}}=\Delta^{\!{\rm tot}}F^{\rm theor}/|F^{\rm theor}|$,
one should combine the errors $\delta_{0}^{\rm theor}$ and
$\delta_{3}^{\rm theor}$. In doing so we take into account that these 
errors are described by a nonuniform  and uniform distributions, 
respectively.
The quantity $\delta^{\!{\rm theor}}$ as
a function of separation is plotted in Fig.~2 as a dashed curve.
Finally, combining the total experimental, 
$\Delta^{\!{\rm tot}}F^{\rm expt}$,
 and theoretical, $\Delta^{\!{\rm tot}}F^{\rm theor}$, errors in 
a conservative way, 
we obtain the resulting absolute error
$\Xi(z)$ for the difference $F^{\rm theor}(z)-F^{\rm expt}(z)$. 

Now we are in a position to compare  theory with the experiment. In Fig.~3
the differences of the theoretical and mean experimental Casimir
forces (shown in Fig.~1) are plotted. In the same figure 
the solid curves exhibit the confidence interval 
$\left[-\Xi(z),\Xi(z)\right]$
computed for any $z$ at 95\% confidence. As is seen from Fig.~3,
almost all differences between the theoretical and
experimental forces (not just 95\% of them as is required by the accepted
confidence) are well within the confidence interval, i.e.,
theory is in excellent agreement with data [we do not plot the results
at $z>425\,$nm as the force magnitudes there are less than
$\Xi(z)$]. 
Quantitatively, the rigorous
measure of agreement between theory and experiment is equal to
$\Xi(z)/|F^{\rm theor}|$. This quantity results in the smallest
value of 3.8\% within the separation region from 75.8\,nm to 81.5\,nm.
It is notable, however, that the actual difference between the theoretical
and experimental force values are less than 1\% of force magnitude
within the separations from 62.33\,nm to 69.98\,nm. At the same time
the rigorous measure of agreement in this interval varies between
4.15\% and 3.9\%.

To conclude, we have performed the first measurement of the Casimir force
between large Au sphere and single crystal Si plate with experimental 
relative error equal to 0.87\% at the shortest separation.
Data are found to be in excellent agreement with theory demonstrating
that this measurement is both precise and accurate. At the same time
the uncertainties in the measurement of surface separations do not
permit one to obtain the theoretical results of the same precision as
the experimental ones at separations less than 100\,nm. 
The case of metal-semiconductor
test bodies appears to be quite different from the case of dissimilar
metals Au-Cu \cite{13} where no noticeable changes of the force magnitude
were found in comparison with the Au-Au system. Here the ratio of the
Casimir forces between Au and Si to Au-Au is 0.74 at the shortest
separation. At a separation of 200\,nm it is only 0.63.
This reduction can be understood physically from lower reflectivity
of a semiconductor in comparison to a metal. The distance
dependence of the above ratio is explained by the fact that the
force between Au-Au bodies decreases with the increase of
separation distance more slowly than between Au-Si bodies.
We note that if our silicon plate were to behave as a metal
instead of a semiconductor, the ratio under discussion
would be practically constant or increase with the increase of 
separation. This qualitatively new
behavior of a metal-semiconductor system in comparison with the case of
two metals opens new opportunities for the modulation of the Casimir
force due to material properties in micro- and nanoelectromechanical
systems.

This work was supported by the NSF Grant PHY0355092 and
DOE grant DE-FG02-04ER46131. G.L.K. and V.M.M. were also
supported by Finep.



\begin{figure}
\vspace*{-7cm}
\centerline{
\includegraphics{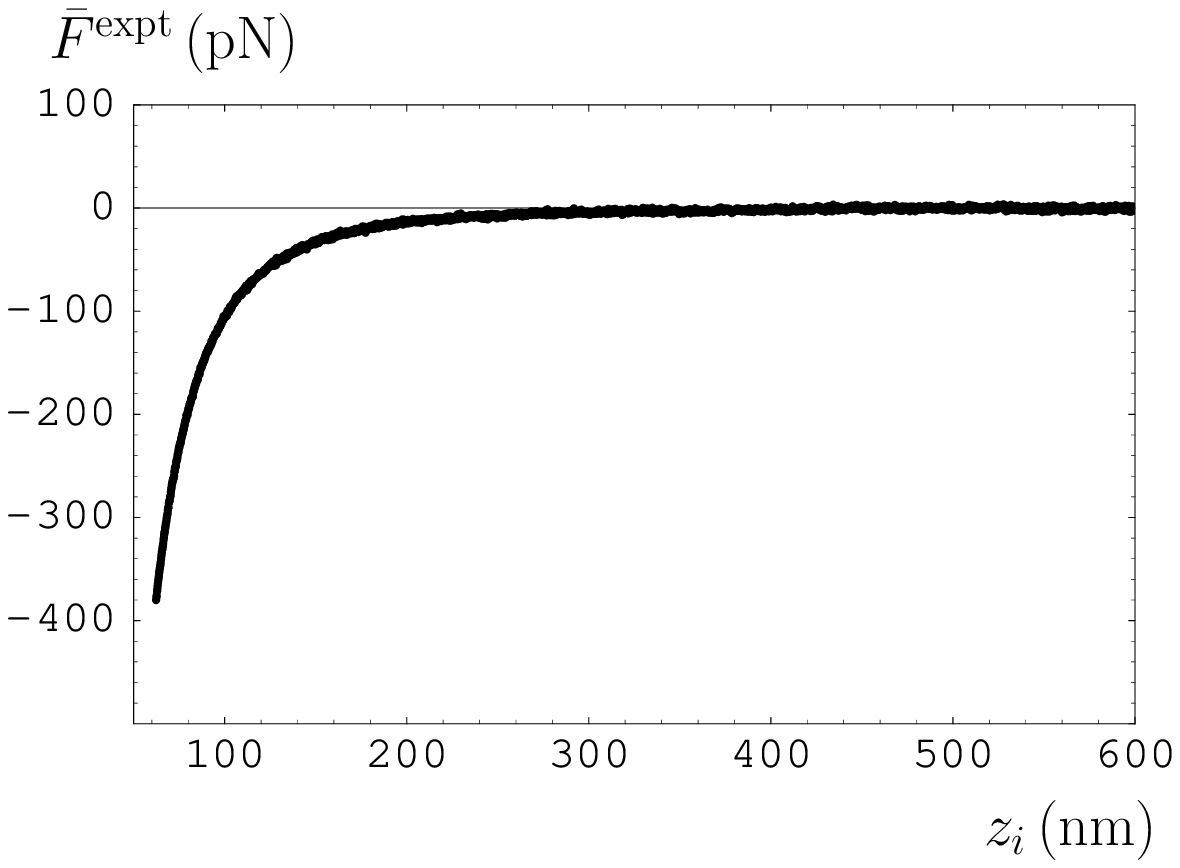}
}
\vspace*{-6cm}
\caption{The mean measured Casimir
force  as a function of separation between Si plate and Au sphere. }
\end{figure}

\begin{figure}
\vspace*{-7cm}
\centerline{
\includegraphics{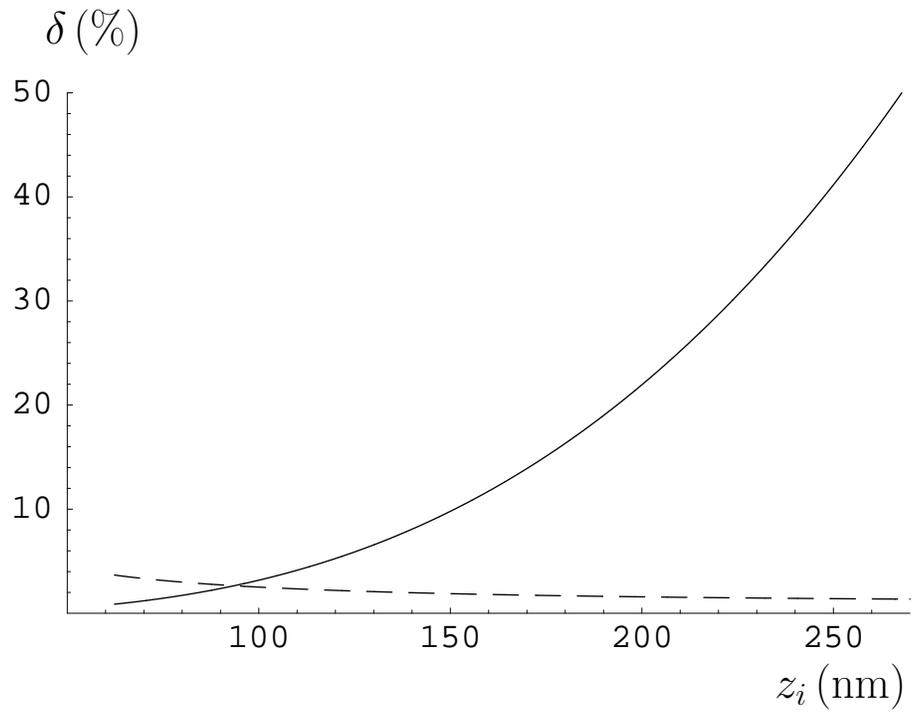}
}
\vspace*{-6cm}
\caption{The total relative experimental $\delta^{\rm expt}$
(solid curve) and theoretical $\delta^{\rm theor}$
(dashed curve) errors as a function of 
plate-sphere separation.}
\end{figure}
\begin{figure}
\vspace*{-7cm}
\centerline{
\includegraphics{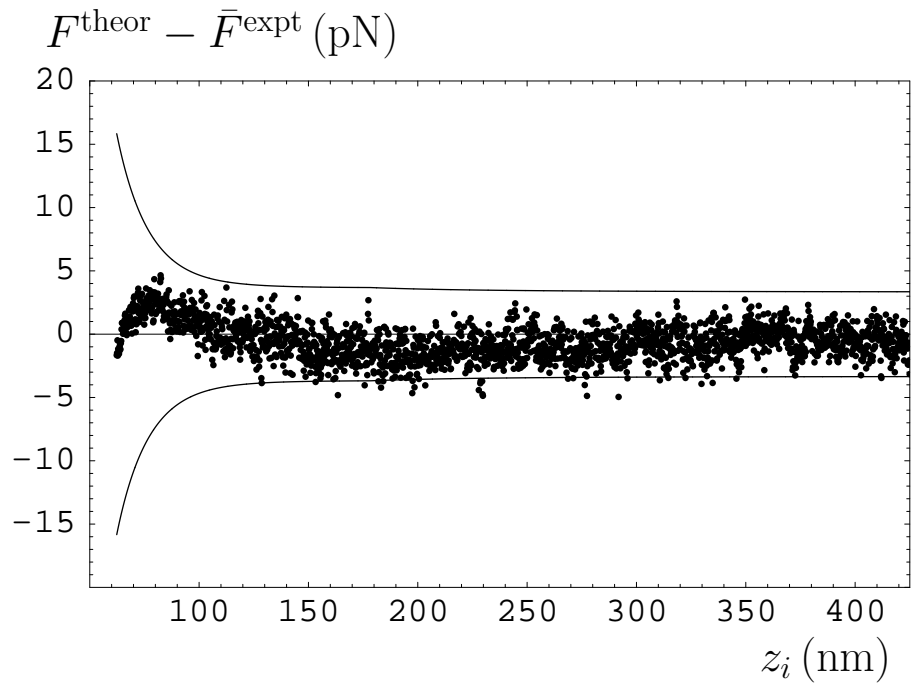}
}
\vspace*{-6cm}
\caption{The 95\% confidence intervals (solid curves) and differences
between theoretical and mean measured Casimir forces versus
plate-sphere separation.}
\end{figure}
\end{document}